\documentclass[twocolumn, 11pt, a4paper]{article}
\usepackage[absolute]{textpos}
\usepackage[top=1.5cm,bottom=3cm,left=1.6cm,right=1.6cm]{geometry}
\usepackage{amsmath,amssymb,mathtools,dsfont,emptypage,graphicx,dcolumn,bm,booktabs,colortbl,collcell,enumerate,float,verbatim,multirow,xparse,slashed,mathrsfs,color}
\usepackage{cite}
\usepackage{tabularx}
\usepackage{multirow}
\usepackage{graphicx}
\usepackage{subcaption}
\usepackage{appendix}
\usepackage{booktabs} 
\usepackage{colortbl}
\usepackage{collcell}
\usepackage{xparse}
\usepackage{relsize}
\usepackage{multicol}
\usepackage{cuted}
\usepackage{titlesec}
\usepackage{float}
\usepackage{ragged2e}
\usepackage[labelfont=bf, labelsep=period]{caption}
\usepackage[compat=1.1.0]{tikz-feynman}
\usepackage{contour}
\usepackage{footmisc}
\usepackage{scrextend}
\deffootnote[1em]{1.5em}{1em}{\textsuperscript{\thefootnotemark}\,}
\usepackage{etoolbox} 
\titleformat{\section}
  {\normalfont\large\bfseries} 
  {\thesection. }{.1em}{}         

\titleformat{\subsection}
  {\normalfont\normalsize\bfseries}
  {\thesubsection. }{.1em}{}

\titleformat{\subsubsection}
  {\normalfont\normalsize\bfseries}
  {\thesubsubsection. }{.1em}{}

\BeforeBeginEnvironment{pmatrix}{
  \setlength{\arraycolsep}{1.5pt} 
}
\AfterEndEnvironment{pmatrix}{
  \setlength{\arraycolsep}{5pt} 
}

\usepackage[compat=1.1.0]{tikz-feynman}
\usepackage{contour}

\def\LNV{\text{LNV}}
\def\CKM{\mathsmaller{\rm CKM}}

\def\min{\mathsmaller{\rm min}}

\newcommand{\Acp}{A_\text{CP}}

\definecolor{myred}{rgb}{.9, 0.1, 0.2}
\definecolor{myblue}{rgb}{0.27, 0.5, 0.9}
\definecolor{mygray}{gray}{.95}
\usepackage[colorlinks=true,urlcolor=myred,linkcolor=myred,citecolor=myred]{hyperref}

\begin{document}

\twocolumn[{%
\begin{@twocolumnfalse}
\begin{flushright}
UdeM-GPP-TH-26-313
\end{flushright}

\begin{center}
{\bf\boldmath \Large Testing Leptogenesis via CP asymmetries\\ in
Lepton-Number-Violating $W$ Decays at Colliders}\\
[2mm]
\renewcommand\*{\thefootnote}{\fnsymbol{footnote}}
Jean-François Fortin$^{\,a,}$\hyperlink{email1}{${}^\dagger$},
Alessandro Granelli$^{\,b,}$\hyperlink{email2}{${}^\ddagger$},
David London$^{\,c,}$\hyperlink{email3}{${}^\S$}, \\
Serguey T. Petcov$^{\,d,e,}$\hyperlink{email4}{${}^\parallel$}
and
Nicolas Sanfaçon$^{\,c,}$\hyperlink{email5}{${}^\perp$}
\\
\vspace{1mm}
$^{a}${\it D\'epartement de Physique, de G\'enie Physique et d'Optique,
Universit\'e Laval,\\ Qu\'ebec, QC, Canada G1V 0A6}\\
$^{b}${\it Instituto de Física Corpuscular (IFIC), CSIC-Universitat de València, Parc Científic UV,\\
c/ Catedrático José Beltrán 2, E-46980 Paterna, Spain}\\
$^{c}${\it Physique des Particules, Universit\'e de Montr\'eal,
1375, ave Th\'er\`ese-Lavoie-Roux,\\ Montr\'eal, QC, Canada H2V 0B3}\\
$^{d}${\it INFN/SISSA, via Bonomea 265, 34136, Trieste, Italy}\\
$^{e}${\it Kavli IPMU (WPI), UTIAS, University of Tokyo, Kashiwa, Chiba 277-8583, Japan}
\end{center}

In principle, searches for CP asymmetries in rare $W$-boson decays can test low-scale leptogenesis in the GeV-scale type-I seesaw with three quasi-degenerate-in-mass heavy right-handed neutrinos. To investigate this, we focus on the lepton-number-violating heavy-neutrino-mediated decay $W^{\pm}\to e^{\pm} e^{\pm} (\bar{q} q')^{\mp}$, with $(\bar{q} q')^+ = (\bar{d}u)^+$ and $(\bar{q} q')^- = (\bar{u}d)^-$. We determine the minimum branching ratio required for a CP asymmetry to be observable at colliders with a given statistical significance. We then perform a numerical scan of the presently-allowed parameter space, also solving the density matrix equations for low-scale leptogenesis, and identify regions where, simultaneously, the present baryon asymmetry of the Universe can be achieved and the collider CP asymmetry for the considered decay is observable at HL-LHC and FCC-hh. We find that these regions are sizeable. A future observation of a CP asymmetry in $W^{\pm}\to e^{\pm} e^{\pm} (\bar{q} q')^{\mp}$ within the regions identified here would provide evidence in favour of low-scale leptogenesis as the origin of the matter-antimatter asymmetry of the Universe.\\
\hrule
\vspace{.5em}
\end{@twocolumnfalse}
}]

\begingroup
\renewcommand{\thefootnote}{}
\footnotetext{%
\hspace{.45em}\hypertarget{email1}{${}^\dagger$}
\href{mailto:jean-francois.fortin@phy.ulaval.ca}{jean-francois.fortin@phy.ulaval.ca}\\
\hypertarget{email2}{${}^\ddagger$}
\href{mailto:alessandro.granelli@ific.uv.es}{alessandro.granelli@ific.uv.es}\\
\hypertarget{email3}{${}^\S$}
\href{mailto:london@lps.umontreal.ca}{london@lps.umontreal.ca}\\
\hypertarget{email4}{${}^\parallel$}
\href{mailto:petcov@sissa.it}{petcov@sissa.it};
also at Institute of Nuclear Research and Nuclear Energy, Bulgarian Academy of Sciences, 1784 Sofia, Bulgaria.\\
\hypertarget{email5}{${}^\perp$}
\href{mailto:nicolas.sanfacon@umontreal.ca}{nicolas.sanfacon@umontreal.ca}
}
\endgroup

\renewcommand*{\thefootnote}{\arabic{footnote}}
\setcounter{footnote}{0}

\noindent {\bf \textit{Introduction.}}~The origin of the matter-antimatter or baryon asymmetry of the Universe, $\eta_B \equiv (n_b - n_{\bar b})/n_\gamma \neq 0$, $n_b$, $n_{\bar b}$ and $n_\gamma$ being the present baryon, antibaryon and photon number densities, remains one of the foremost open questions in particle physics and cosmology. Data on the cosmic microwave background anisotropies \cite{Planck2018} and primordial light-element abundances \cite{Cooke_2018} yield $\eta_B^{\rm obs}\simeq 6.1\times10^{-10}$. 

A compelling explanation is that $\eta_B^{\rm obs}$ originated from the same physics beyond the Standard Model (SM) responsible for generating light neutrino masses, another long-standing challenge in the field. Here the simplest SM extension  is the type-I seesaw \cite{Minkowski:1977sc,Yanagida:1979as,Gell-Mann:1979ijt,GellMann:1980vs,Glashow:1979nm,Mohapatra:1979ia}. This involves adding to the SM $n_R\geq 2$ singlet right-handed (RH) neutrinos which have Majorana masses and couple to the three left-handed (LH) lepton doublets via Yukawa interactions. After electroweak symmetry breaking, the Yukawa coupling generates neutrino Dirac mass terms that connect the LH and RH neutrino fields. Together with the RH neutrino Majorana masses, these form the full neutrino mass matrix. Upon its diagonalisation, one obtains three light (sub-eV scale) and, in general, $n_R$ heavy (much heavier than the eV scale) Majorana neutrinos, respectively $\nu_a$ and $N_j$, with masses $m_a$ and $M_j\gg m_a$, $a=1,2,3$, $j =1,...,n_R$. The LH flavour and the RH neutrinos are linear combinations of these mass eigenstates. The mixing matrix between the flavour neutrinos and the heavy states $N_j$, $\Theta$, contains CP-violating phases. As a consequence of this mixing, the $N_j$ have 
couplings to the SM
leptons 
and 
gauge bosons.

The type-I seesaw and the expansion of the Universe provide the necessary Sakharov's conditions \cite{Sakharov:1967dj} to generate a primordial lepton asymmetry through {\it leptogenesis} \cite{Fukugita:1986hr,Akhmedov:1998qx} (see \cite{Bodeker:2020ghk} for a review), thanks to the CP- and lepton-number-violating (LNV) Yukawa interactions involving the RH neutrinos (decays, scatterings), and their oscillations. The lepton asymmetry is then converted into $\eta_B$ by sphalerons \cite{Sakharov:1967dj,Klinkhamer:1984di,Kuzmin:1985mm, DOnofrio:2014rug}. The resulting value of $\eta_B$ is a very complicated function of the heavy neutrino masses, their Yukawa couplings, and the CP-violating phases contained in these couplings.

Leptogenesis can only be tested indirectly, through the phenomenology of the particle-physics models that realise it.
The low-scale realisation of the type-I seesaw \cite{Shaposhnikov:2006nn,Kersten:2007vk,Gavela:2009cd} (see also \cite{Wyler:1982dd,Leung:1983ti,Mohapatra:1986aw,Malinsky:2005bi}) and the corresponding leptogenesis scenarios \cite{Akhmedov:1998qx,Pilaftsis_1992,Pilaftsis:1997jf,Klaric:2020phc,Klaric:2021cpi} are particularly promising in this regard. They rely on quasi-degenerate-in-mass heavy neutrinos at the accessible GeV-to-TeV scales, with an associated experimentally testable phenomenology \cite{delAguila:2007qnc,Gorbunov:2007ak,delAguila:2008cj,delAguila:2008hw,Atre:2009rg,Ibarra:2010xw,Ibarra:2011xn,Dinh:2012bp,Cely:2012bz,Antusch:2015mia, Penedo:2017knr,Bondarenko:2018ptm,Bolton:2019pcu,Granelli:2022eru,Urquia-Calderon:2022ufc, Granelli:2025lds, Abdullahi:2022jlv,Antel:2023hkf}. 

Within this phenomenology, the possibility of measuring CP violation in LNV heavy-neutrino-mediated decays of mesons \cite{Cvetic:2013eza, Cvetic:2014nla, Dib:2014pga, Cvetic:2015naa, Cvetic:2015ura, Cvetic:2020lyh, Godbole:2020jqw, Zhang:2020hwj, Zhang:2022zpi}, the $\tau$ lepton \cite{Zamora-Saa:2016ito, Tapia:2019coy, Cvetic:2023bwr}, the $W$ boson \cite{Najafi:2020dkp, Datta:1993nm, Ali:2001gsa, Han:2006ip, Chen:2013foz, Deppisch:2015qwa, Izaguirre:2015pga, Degrande:2016aje, Das:2016hof, Antusch:2016ejd, Das:2017nvm, Cai:2017mow, Lu:2022pvw, Abada:2022wvh, Lu:2022wsm, Antusch:2026hwc}, the top quark \cite{Bar-Shalom:2006osy, Si:2008jd, Eilam:2008qtl, Liu:2019qfa, Lu:2021vzj}, as well as scattering processes \cite{Alva:2014gxa, Ruiz:2017nip, Fuks:2020att, Wang:2023zhh} has been studied. At the same time, testing low-scale type-I seesaw and leptogenesis via heavy-neutrino searches at colliders has also been extensively explored \cite{Canetti:2014dka,Hernandez:2016kel,Drewes:2016jae,Antusch:2017pkq,Abada:2018oly,Drewes:2021nqr,Drewes:2024bla}. 

However, even if a CP asymmetry at colliders were observed, it would not be clear whether this measurement is compatible with leptogenesis. Answering this question is extremely non-trivial, since collider CP asymmetries and $\eta_B$ depend on the same seesaw parameters, but in very different ways. The region of the parameter space that produces $\eta_B=\eta_B^{\rm obs}$ may or may not overlap with the one that yields the measured CP asymmetry.

In this paper, we analyse the simultaneous realisation of an observable CP asymmetry in the LNV decays $W^\pm\to e^\pm e^\pm(\bar qq')^\mp$, with $(\bar qq')^+=(\bar du)^+$ and $(\bar qq')^-=(\bar ud)^-$, in which the $W^\pm$ is produced on-shell at hadron colliders, and successful GeV-scale leptogenesis with three quasi-degenerate heavy neutrinos. To be specific, by performing a scan of the parameter space and computing the considered $W$ decay rate, CP asymmetry and $\eta_B$ at each point, we find a sizeable common region in which (i) the rate and CP asymmetry of the considered $W$ decay are potentially measurable at HL-LHC \cite{ZurbanoFernandez:2020cco} and FCC-hh \cite{FCC:2018vvp,FCC:2019,FCC:2025lpp}, and (ii) $\eta_B^{\rm obs}$ is reproduced. To our  knowledge, this is the first joint study of this type.\\

\noindent{\bf \textit{Branching ratio and  CP asymmetry.}}~We consider the LNV decay $W^\pm \to \ell_1^\pm \ell_2^\pm (\bar q q')^\mp$, with $\ell_{1,2}$ being charged leptons and $q, q'$ quarks. The decay is mediated by the on-shell heavy neutrinos $N_j$. In the narrow-width approximation, neglecting interference terms in the total rate, its branching ratio can be written as
\begin{equation}
\mathcal B
\simeq
\sum_{j=1}^{n_R}
\frac{\Gamma(W^\pm\to\ell_1^\pm N_j)}{\Gamma_W}
\frac{\Gamma(N_j\to\ell_2^\pm(\bar q q')^\mp)}{\Gamma_j},
\label{eq}
\end{equation}
where $\Gamma_W \simeq 2.085\,\text{GeV}$ \cite{ParticleDataGroup:2024cfk} is the total decay width of the $W$ boson, and $\Gamma_j$ that of the heavy neutrino $N_j$. $\Gamma_j$ is obtained by summing all kinematically-allowed charged- and neutral-current decay channels of $N_j$. We consider $M_j\lesssim M_W \simeq 80.4\,\text{GeV}$, so these decays proceed through off-shell $W^\pm$ and $Z$ bosons; Higgs-mediated contributions are negligible in the mass range of interest. The hadronic final states are treated in the free-quark approximation, including QCD corrections, following~\cite{Bondarenko:2018ptm}. We consider $M_j\geq 1\,\text{GeV}$ so that this perturbative treatment is applicable, while the residual uncertainty from the hadronic description has negligible impact.

The partial decay widths $\Gamma(W^\pm\to\ell_1^\pm N_j) = G_F |\Theta_{\ell_1j}|^2 F_W$ and $\Gamma(N_j\to\ell_2^\pm(\bar q q')^\mp) = G_F^2 |V^{\CKM}_{q'q}|^2 |\Theta_{\ell_2 j}|^2 F_N$ are computed using the standard two- and three-body decay expressions, respectively, retaining the full mass and width dependence of the particles involved. Here, $F_W$ and $F_N$ encode the corresponding mass and phase-space dependence, $G_F$ is the Fermi constant, $V^{\CKM}_{q'q}$ is the relevant Cabibbo-Kobayashi-Maskawa (CKM) matrix element and $\Theta_{\ell j}$ is the mixing of the heavy neutrino $N_j$ with the active neutrino of flavour $\ell = e,\,\mu,\,\tau$. The full expressions are given in a companion paper \cite{PRD_version}. 

Our calculation for the CP asymmetry $\Acp^{\LNV} \equiv (\Gamma^--\Gamma^+)/(\Gamma^-+\Gamma^-)$, with $\Gamma^\pm \equiv \Gamma(W^\pm \to \ell_1^\pm \ell_2^\pm (\bar{q} q')^\mp)$ follows~\cite{Najafi:2020dkp}; details are given in \cite{PRD_version}. In the quasi-degenerate limit of $M_N\equiv M_j \simeq M_k$ and $\Gamma_N\equiv \Gamma_j \simeq \Gamma_k$, $j,k=1,...,n_R$,  we find that its characteristic dependence reads 
\begin{align}
\begin{split}
    \Acp^{\LNV} \propto& \sum_{j=2}^{n_R}\sum_{k<j} \frac{2M_W \Gamma_N^2 \delta M_{jk}r\left(r+2 \right)}{[r^2\delta M_{jk}^2 + \Gamma_N^2] [\delta M_{jk}^2 + \Gamma_N^2]}\times \\
    &\hspace{20mm}\times\,\Im[\Theta_{\ell_1 k}^* \Theta_{\ell_1 j} \Theta_{\ell_2 k}^* \Theta_{\ell_2 j}],
    \end{split}
\end{align}
where $\delta M_{jk}\equiv M_j-M_k$ and $r \equiv M_N/M_W$.
For $r \leq 1$, each term in the sum is maximised for a mass splitting close to the decay width, $\delta M_{jk} \sim \Gamma_N$. When this Breit-Wigner-like resonance condition is satisfied for a given heavy neutrino pair, the CP asymmetry is enhanced.

This resonance condition -- which differs from the one relevant for resonant leptogenesis \cite{Pilaftsis_1992,Pilaftsis:1997jf}, as the latter occurs before electroweak symmetry breaking -- is instrumental for obtaining an observable CP asymmetry at colliders. It also has important consequences for low-scale leptogenesis. For $n_R=2$, the leading CP-violating sources of GeV-scale leptogenesis are described by the basis-invariant structures identified in \cite{Hernandez:2022ivz} and are proportional to $\delta M_{21}$ at leading order. The collider resonance requires an extremely small mass splitting that, in the mass range $1\lesssim M_N/\text{GeV}\lesssim 80$ of interest, needs to vary within $10^{-20} \,(10^{-18}) \lesssim \delta M_{21}/M_N \lesssim 10^{-12}\,(10^{-10})$ for $\Theta^2 \equiv \sum_{\ell, j} |\Theta_{\ell j}|^2 \sim 10^{-7}\,(10^{-5})$, strongly suppressing these leading order leptogenesis invariants. Higher-order invariants can remain non-vanishing in the degenerate limit, but at the cost of a reduced parameter space  \cite{Sandner:2023tcg}.

The inclusion of a third heavy neutrino $N_3$ provides additional freedom to alleviate the suppression in $\eta_B$. The third heavy neutrino allows further flexibility to reproduce the observed light-neutrino mass spectrum and, during leptogenesis, can be separated from the pseudo-Dirac pair formed by the first two heavy neutrinos and equilibrate at a later stage \cite{Georis:2024qgk}. The corresponding asymmetry can remain comparatively protected from washout effects and the parameter space of viable leptogenesis enlarges considerably \cite{Abada:2018oly, Drewes:2021nqr}, while the collider phenomenology remains qualitatively unaltered. We therefore consider $n_R=3$ in our analysis, choosing $\delta M_{21} \sim \Gamma_N$ to allow for a resonant enhancement of the CP asymmetry and a sufficiently larger splitting $\delta M_{31}$ to ensure efficient generation of the baryon asymmetry during leptogenesis. \\

\noindent{\bf \textit{Collider detectability.}}~The observability of the LNV $W$ decays is limited by the finite lifetime of the heavy neutrinos, as they have to travel from the interaction point and decay inside the detector. The probability for a heavy neutrino $N_j$ with momentum $\bm{p}_j$ to decay within a distance $L$ from the interaction point is
\begin{equation}
P_j(L)=1-\exp\!\left[-\frac{M_j\Gamma_j L}{|\bm{p}_j|}\right].
\end{equation}
To take this probability into account, we define the effective branching ratio as $\mathcal{B}_{\rm eff}=\sum_j P_j(L)\mathcal{B}_j$, where $\mathcal{B}_j$ denotes the contribution of $N_j$ to the total $\mathcal{B}$. We adopt $L=1\,\mathrm{m}$, representative of the inner tracking volume of the LHC detectors \cite{ATLAS:2023dns,CMS:2023gfb}.

For an $n\sigma$ measurement of a CP asymmetry $A_{\rm CP}$, the minimum measurable branching ratio is given by \cite{Najafi:2020dkp}
\begin{equation}
\mathcal{B}_{\rm min}
=
\frac{n^2(1-A_{\rm CP}^2)}
{\epsilon A_{\rm CP}^2\,\mathcal{L}_{\rm int}\sigma_{pp\to WX}},
\end{equation}
where $\mathcal{L}_{\rm int}$ is the integrated luminosity and $\sigma_{pp\to WX}$ is the inclusive $W$-boson production cross section, with $X$ denoting the accompanying hadronic system. Taking $n=3$ and $\epsilon=1\%$, we obtain $\mathcal{B}_{\rm min}\simeq3.3\times10^{-7}$ and $1.5\times10^{-8}$ for $A_{\rm CP}=30\%$ at LHC Run~2 ($\mathcal{L}_{\rm int} = 147\,\text{fb}^{-1}$, $\sigma_{pp\to WX} = 190\,\text{nb}$ \cite{ATLAS:2023dns, ATLAS:2016fij, ParticleDataGroup:2024cfk}) and HL-LHC ($\mathcal{L}_{\rm int} = 3\,\text{ab}^{-1}$, $\sigma_{pp\to WX} = 200\,\text{nb}$ \cite{ATLAS:2019mfr, Hou:2019efy}), respectively, while FCC-hh ($\mathcal{L}_{\rm int} = 30\, \text{ab}^{-1}$, $\sigma_{pp\to WX} = 1.3 \,\text{\textmu b}$ \cite{FCC:2018vvp, FCC:2019, FCC:2025lpp}) could reach $\mathcal{B}_{\rm min}\simeq2.3\times10^{-10}$. For the more conservative $A_{\rm CP}=10\%$, these values increase by approximately one order of magnitude.\\ 

\noindent{\bf \textit{GeV-scale leptogenesis and the joint analysis.}} The dynamics of low-scale leptogenesis with quasi-degenerate heavy Majorana neutrinos at the GeV mass scale \cite{Akhmedov:1998qx,Klaric:2020phc,Klaric:2021cpi} can be described by a set of momentum-averaged density-matrix equations \cite{Drewes:2016gmt,Hernandez:2016kel,Ghiglieri:2017gjz, Eijima:2017anv, Antusch:2017pkq, Abada:2018oly}. These equations describe the production and coherent CP-violating oscillations of the heavy-neutrino states, together with the generation and washout of the individual lepton-flavour asymmetries through decays, inverse decays, and scatterings involving the heavy neutrinos, Higgs boson and SM leptons. Thermal and spectator effects are consistently included. We solve these equations numerically using the \texttt{ULYSSES} Python package \cite{Granelli:2020pim,Granelli:2023vcm,Granelli:2026goh} and compute $\eta_B$ in terms of the same seesaw parameters that also control $\mathcal{B}_{\rm eff}$ and $A_{\rm CP}^{\LNV}$. We refer to the latest \texttt{ULYSSES} manual \cite{Granelli:2026goh} and the companion paper \cite{PRD_version} for the explicit equations and their numerical implementation. 

The heavy neutrino mixing matrix $\Theta$ parametrises the coupling of heavy neutrinos to the SM weak currents, governing their production and decays at colliders. It is related to the Majorana mass matrix of the RH neutrinos and their Yukawa couplings to the Higgs and charged-lepton doublets, which in turns regulate the light neutrino masses and mixing, and the generation of the lepton asymmetry during leptogenesis. For the purpose of our scan, we parametrise it using the Casas-Ibarra (CI) construction \cite{Casas:2001sr}, including the one-loop radiative corrections to the light-neutrino mass matrix \cite{Lopez-Pavon:2015cga}: $\Theta = \pm i\,U\,\sqrt{\hat m_\nu}\,O\sqrt{M_{\rm loop}}\,\hat M_N^{-1}$, where $\hat m_\nu=\mathrm{diag}(m_1,m_2,m_3)$ is the diagonal mass matrix for the active neutrinos; $M_{\rm loop}$ is the modified heavy-neutrino mass matrix accounting for the one-loop corrections to active neutrino masses \cite{Pilaftsis_1992, Grimus_2002, Aristizabal_Sierra_2011, Lopez_Pavon_2013}; $U$ is the Pontecorvo-Maki-Nakagawa-Sakata (PMNS) neutrino mixing  matrix \cite{Pontecorvo:1957cp, Pontecorvo:1957qd, Pontecorvo:1967fh, Maki:1962mu}; $O$ is a complex orthogonal matrix. For $O$ we adopt the parametrisation of~\cite{Drewes:2021nqr} (see also \cite{Granelli:2026qss}): $O=R^{(13)}(\omega^\nu_2)R^{(23)}(\omega^\nu_1)R^{(12)}(\omega + i \xi)R^{(23)}(\omega^N_1)R^{(13)}(\omega^N_2)$, where $R^{(ij)}$ denotes a $3\times3$ orthogonal rotation in the $(i,j)$ plane, with $\omega,\xi,\omega^\nu_1,\,\omega^\nu_2,\,\omega^N_1,\,\omega^N_2$ real. We then consider the standard PMNS parametrisation \cite{ParticleDataGroup:2024cfk}, but with the two Majorana phases set to zero for simplicity. The three mixing angles, the two mass-squared differences and the Dirac phase are fixed to their \texttt{NuFit~6.0} best-fit values \cite{Esteban:2024eli}. We show the results in the case of normal ordering, $m_1<m_2<m_3$. The results for inverted ordering, $m_3<m_1<m_2$ are similar and are provided in \cite{PRD_version}.

As is well-known, in the type-I seesaw the heavy-neutrino mass eigenstates $N_j$, with masses $M_j$, after electroweak symmetry breaking generally differ from those relevant for leptogenesis in the electroweak-symmetric phase, where most of the mechanism takes place. We account for this difference by reconstructing the RH neutrino mass matrix at second order in the seesaw expansion, $M_R\simeq \hat M_N-(1/2)\Theta^\dagger\Theta\hat M_N -(1/2)\hat M_N\Theta^T\Theta^*$, where $\hat M_N=\mathrm{diag}(M_1,M_2,M_3)$. We then diagonalise $M_R=\tilde V\tilde M_N\tilde V^T$ to obtain the heavy neutrino masses in the symmetric phase and calculate the Yukawa couplings $Y=(\sqrt{2}/v)\tilde\Theta M_N\tilde V^*$, with $v=246\,\mathrm{GeV}$, as relevant for leptogenesis calculations.

We scan over the mass range $1\leq M_1/\mathrm{GeV}\leq80$, varying the mass splitting $\delta M_{31}$ and the other CI parameters, and imposing $\Delta M_{21}\sim\Gamma_N$ to enhance the collider CP asymmetry. We consider $m_1=0,\,0.05,\,0.1\,\mathrm{eV}$ and choose the CI parameters so that $\Theta_\tau^2 \ll \Theta^2_\mu,\,\Theta^2_e$ with $\Theta^2_\ell \equiv \sum_{j=1}^3 |\Theta_{\ell j}|^2$, $\ell = e,\,\mu,\,\tau$, suppressing wash-outs in the tau flavour and favouring successful leptogenesis at large mixings~\cite{Abada:2018oly,Hernandez:2022ivz}. The scan ranges are $5\times10^{-10}\leq\Delta M_{31}/M_1\leq10^{-2}$, $0.4\leq\omega^N_{1,2}\leq0.8$, $0\leq\omega\leq2\pi$, and $1.4\leq\xi\leq7$ ($10.8$) for $M_1 \leq 20\,\mathrm{GeV}$ ($M_1 > 20\,\mathrm{GeV}$), these ranges being chosen \textit{a posteriori} to maximise the overlap between collider detectability and the viability of low-scale leptogenesis.

\begin{figure}[t!]
    \centering
        \includegraphics[width=\linewidth]{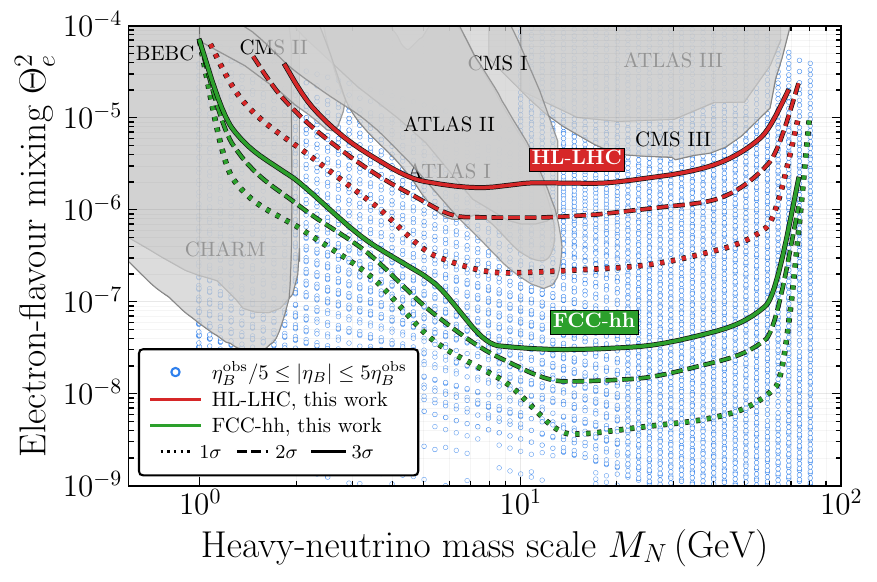}
    \caption{Results of the joint scan of leptogenesis and CP asymmetry in the LNV decay $W^\pm \to e^\pm e^\pm (\bar u d)^-/(\bar d u)^+$, in the
$\Theta_e^2$--$M_N$. The grey regions correspond to constraints from BEBC~\cite{Barouki:2022bkt}, CHARM \cite{CHARM:1985nku}, ATLAS (I \cite{ATLAS:2022atq}, II \cite{ ATLAS:2025uah}, III \cite{ATLAS:2025lva}) and CMS (I \cite{CMS:2022fut}, II \cite{CMS:2024ake}, III \cite{CMS:2024xdq}). Note that the projected bounds assume electron-flavour-dominated mixing and either one or two heavy neutrinos, and should therefore be regarded as indicative in our scenario~\cite{Tastet:2021ygq,Abada:2022wvh}. The blue points are those of our scan that satisfy our successful leptogenesis criterium. The red (green) curves obtained in this work denote the minimum $\Theta_e^2$ required for the CP asymmetry in the considered process to be probed at the HL-LHC (FCC-hh), while maintaining successful leptogenesis. Solid, dashed and dotted lines correspond to $3\sigma$, $2\sigma$ and $1\sigma$ sensitivity, respectively.}
    \label{fig:scanresults}
\end{figure}

For each point, we compute $\mathcal B_{\rm eff}$, $A_{\rm CP}^{\rm LNV}$ in $W^\pm\to e^\pm e^\pm (\bar u d)^-/(\bar d u)^+$, and $\eta_B$. We retain points satisfying $0.2\,\eta_B^{\rm obs}\leq|\eta_B|\leq5\,\eta_B^{\rm obs}$, since the observed asymmetry can subsequently be reproduced by adjusting the magnitude and sign of $\Delta M_{31}$ without affecting the collider phenomenology~\cite{PRD_version}. Among these points, we identify those that yield an effective branching ratio above $\mathcal{B}_{\min}$ for HL-LHC and FCC-hh and thus derive the corresponding minimum heavy neutrino mixing required to have both successful leptogenesis and an observable CP asymmetry at colliders.\\

\noindent {\bf \textit{Results of the scan and conclusion.}} ~Figure~\ref{fig:scanresults} reveals the potential discovery reach of the HL-LHC and FCC-hh in the $\Theta_e^2$--$M_N$ plane. Crucially, it demonstrates that future colliders can measure the CP asymmetry of the considered $W$ decay in regions of the parameter space where $\eta_B^{\rm obs}$ can be reproduced. In the figure, the red and green curves show for each collider the minimum values of $\Theta_e^2$ for which the requirements of successful leptogenesis and of the effective branching ratio of $W^\pm\to e^\pm e^\pm(\bar u d)^-/(\bar d u)^+$ are simultaneously satisfied. 

For the first time, our analysis uncovers broad, uncharted regions of the GeV-scale type-I seesaw parameter space where three quasi-degenerate-in-mass heavy neutrinos can yield both the observed baryon asymmetry $\eta_B^{\rm obs}$ and measurable CP asymmetries in LNV $W$ decays for the same set of model parameters.

Moreover, these conclusions directly extend to other flavour channels. Final states that involve CKM matrix elements $V^{\CKM}_{q'q}$ and flavour mixing $\Theta^2_\ell$ of comparable size to that of $W^\pm\to e^\pm e^\pm(\bar{u} d)^-/(\bar{d} u)^+$, such as $W^\pm\to \mu^\pm \mu^\pm(\bar{c} s)^-/(\bar{s} c)^+$, lead to a similar potential reach at future colliders. Even states that involve quarks with off-diagonal CKM matrix elements could well be  within reach of FCC-hh. Furthermore, we find that modes that include $\tau$ leptons are suppressed due to our choice of the flavour mixing hierarchy $\Theta_\tau^2 \ll \Theta^2_\mu,\,\Theta^2_e$, and that lepton-flavour-violating decays $W^\pm\to \ell_1^\pm \ell_2^\mp (\bar q q')^\pm$ lead to suboptimal CP asymmetries in the quasi-degenerate regime. A more thorough flavour analysis can be found in the companion paper \cite{PRD_version}.

Because the CP asymmetry in $W$ decays and in leptogenesis arise from the same seesaw parameters and CP-violating phases, both phenomena are correlated. Therefore, an experimental observation of a CP asymmetry in LNV $W$ decays would provide compelling evidence in favour of low-scale leptogenesis. The planned HL-LHC and proposed FCC-hh offer a concrete opportunity to search for these signatures, promising to open a new chapter in the quest to understand the origin of the matter–antimatter asymmetry of the Universe.\\

\noindent{\bf \textit{Acknowledgements.}}~We thank F.~Najafi for collaboration in the early stages of this project.  A.G.~acknowledges the use of computational resources from the parallel computing cluster of the \href{https://site.unibo.it/openphysicshub/en}{Open Physics Hub} at the Department of Physics and Astronomy of the University of Bologna, Italy. This work was financially supported by NSERC of Canada (J.-F.F., D.L., N.S.) and by FRQNT of Québec (N.S.), by the Spanish grant PID2023-148162NB-C21, MCIN/AEI/10.13039/501100011033, co-funded by the European Union, FEDER (A.G.), 
by the Italian INFN program on Theoretical Astroparticle Physics and by the World Premier
International Research Center Initiative, WPI Initiative, MEXT, Japan (S.T.P.), and in part by the 
European Union's Horizon Europe programme under Marie Skłodowska-Curie Actions – Staff Exchanges (SE) grant agreement No.~101086085-ASYMMETRY (A.G., S.T.P.). Artificial Intelligence (AI) systems (OpenAI's GPT-5.6 Luna accessed through ChatGPT and GitHub Copilot, Google's Gemini 3.6 Flash) were used under the supervision of the authors to support coding and manuscript writing.


\providecommand{\href}[2]{#2}\begingroup\raggedright\endgroup

\end{document}